\documentclass[prl,twocolumn,superscriptaddress,a4paper]{revtex4}
\usepackage{bm,graphicx,amsmath}
\usepackage{epsfig,color}
\usepackage{graphicx}
\usepackage{bm}
\newcommand{\mycomm}[1]{}
\def\COMMENTS{\renewcommand{\mycomm}[1]{\par\noindent\textcolor{red}{\emph{##1}}}}
\COMMENTS

\begin{document}
\title{Teleportation of massive particles without shared entanglement}
\author{A.~S. Bradley}
\affiliation{Australian Research Council Centre of Excellence for Quantum-Atom Optics} 
\affiliation{School of Physical Sciences, University of Queensland, Brisbane, QLD 4072, Australia}
\author{M.~K. Olsen}
\affiliation{Australian Research Council Centre of Excellence for Quantum-Atom Optics} 
\affiliation{School of Physical Sciences, University of Queensland, Brisbane, QLD 4072, Australia}
\author{S.~A. Haine}
\affiliation{Australian Research Council Centre of Excellence for Quantum-Atom Optics} 
\affiliation{Department of Physics, Australian National University, Canberra, Australia}
\author{J.~J. Hope}
\affiliation{Australian Research Council Centre of Excellence for Quantum-Atom Optics} 
\affiliation{Department of Physics, Australian National University, Canberra, Australia}
\begin{abstract}
We propose a method for quantum state transfer from one atom laser beam to another via an intermediate optical field, using Raman incoupling and outcoupling techniques. Our proposal utilises existing experimental technologies to teleport macroscopic matter waves over potentially large distances without shared entanglement. 
\end{abstract}
\maketitle
The instantaneous, disembodied transport of matter through space is of course, absolutely forbidden by the laws of nature. However, in 1993, a proposal by Bennett {\em et al.\/}~\cite{Bennett} used the term {\em teleporting\/} to describe a scheme which uses quantum entanglement to transfer an unknown discrete variable quantum state between a sender and a receiver, who have since become famous as Alice and Bob. The protocols of this scheme were later extended to the transfer of the quantum state of a system with continuous variables~\cite{vaidman,braunstein}, and have been shown to work experimentally, with fidelities of up to $0.85\pm 0.05$ being achieved for the transfer of an unknown continuous variable quantum state~\cite{fidelidade}.  
While the original demonstrations were with optical states, the techniques have also been applied to ions~\cite{ions} and combinations of atoms and light~\cite{Sherson}. 

In this work we propose and analyse a scheme which allows an atom laser beam to disappear at one location and reappear at another, without the use of shared entanglement between the sender and reciever. 
Our system is related to the experimental transfer of optical information to matter waves realised by Ginsberg {\em et al.\/}~\cite{Ginsberg}, but with two important differences. The first is that our intermediate medium is light rather than atoms, so that transmission over much longer distances should be possible. The second is that we do not rely on the slow-light mechanism, which is possibly the cause of the low transfer efficiencies in the Ginsberg demonstration~\cite{ANU}, but instead use a modification of the Raman atom laser outcoupling scheme~\cite{Raman,Paranjape2003}. Although our scheme is quite distinct from what is normally termed quantum teleportation, we feel that it is closer in spirit to the original fictional concept and so will use the term to describe our system.

What differentiates our scheme from what is usually termed quantum teleportation is that our scheme does not require the sender and receiver to share entangled states. Our scheme avoids this requirement as there is no measurement step involved in sending the information. The degree of entanglement is a limiting factor in the achievable fidelity in traditional quantum teleportation experiments; the generation and distribution of highly entangled states is technically challenging as it usually require highly nonlinear processes and is susceptible to loss. As our scheme is not affected by these factors, it may be possible to achieve a much higher teleportation fidelity than with traditional quantum teleportation. 

\begin{figure}\begin{centering}\includegraphics[width=1\columnwidth]{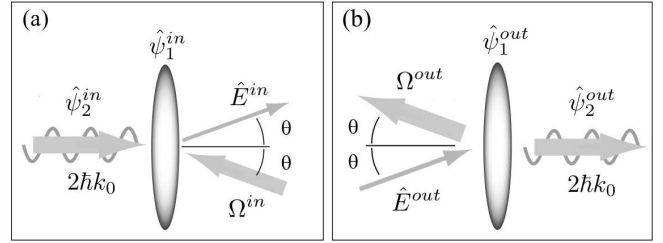}\par\end{centering}
\vspace*{-10pt}
\caption{The sending, $\hat{\psi}_{1}^{in}$, and receiving, $\hat{\psi}_{1}^{out}$, stations for the teleporter. The left panel, (a), shows the {\em sender\/}, which absorbs the propagating atoms and transmits the optical signal containing the quantum information. The right panel, (b), shows the {\em receiver} which absorbs the optical signal and uses the information contained in it to reproduce the original atomic pulse. $\Omega^{in}$ and $\Omega^{out}$ are the intense classical control fields, while $\hat{E}^{in}$ and $\hat{E}^{out}$ are the same probe field, created by the incoupling of $\hat{\psi}_{2}^{in}$ and absorbed by the outcoupling process which creates $\hat{\psi}_{2}^{out}$.}
\label{fig:sendandreceive}
\vspace*{-10pt}
\end{figure}
Our system consists of two separated trapped Bose-Einstein condensates, two classical control optical fields, an input atom laser pulse and a weak optical probe field, as shown schematically in Fig.~\ref{fig:sendandreceive}. 
\begin{figure}[!htb]\begin{centering}\includegraphics[width=.6\columnwidth]{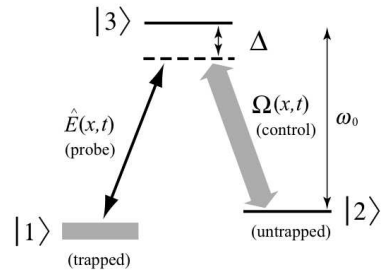}\par\end{centering}
\vspace*{-10pt}
\caption{The electronic level scheme of the atoms involved in the atomic transfer process. The internal states of the atoms are coherently manipulated during incoupling and outcoupling using a two photon Raman transition (see text).}
\label{fig:levels}
\vspace*{-10pt}
\end{figure}
The electronic levels and optical fields used in our scheme are as shown in Fig.~\ref{fig:levels}, with the highly populated levels and fields shown as thick grey lines. This diagram is valid for both the {\em transmitting\/} and {\em receiving\/} condensates, as shown in Fig.~\ref{fig:sendandreceive} (a) and (b). In each case, the weak probe field, represented by $\hat{E}$, plays an important role. 
The sending stage of the scheme is shown in Fig.~\ref{fig:sendandreceive} (a), where the input atom laser beam ($\hat{\psi}_{2}^{in}$) is coupled into the condensate ($\hat{\psi}_{1}^{in}$). The quantum information held in the beam is transferred to the probe field ($\hat{E}^{in}$). The internal Raman energy level configuration allows for stimulated transitions between the trapped and untrapped fields. These transitions are stimulated by both intense optical fields (control), denoted by the Rabi frequencies $\Omega^{in/out}(x,t)$, and highly occupied trapped bosonic matter fields ($\hat{\psi}^{in/out}_{1}$). The optical probe field propagates some distance between the two trapped condensates, and is used ($\hat{E}^{out}$) to outcouple the atom laser field from the second condensate ($\hat{\psi}_{2}^{out}$). The preparation of the sending and receiving condensates and the control lasers needs only the passing of classical information, and can be done without knowing the quantum state of the input beam. State transfer from the input atom laser pulse to the probe field happens automatically given the appropriate conditions, as shown by Bradley {\em et al.\/}~\cite{homodyne}. At the receiving station, the state of the probe field is transferred to the output atomic pulse, using the mechanism described by Haine and Hope~\cite{Haine2005}. If the required conditions are met, there will be a complete transfer of the quantum information contained in the first pulse, via the probe field, to the second. We note that technically our scheme would realise a very efficient quantum channel~\cite{Gu,Shor} to transfer information between condensates.

We perform the mathematical analysis of the system using a one-dimensional model and a computational technique developed by Haine and Hope~\cite{Haine2005}. The {\em sender\/} and {\em receiver\/} can be treated as a cascaded system, so we begin with the {\em sender\/} of Fig.~\ref{fig:sendandreceive}(a), described by the Hamiltonian 
${\cal H}^{in}={\cal H}^{in}_{atom}+{\cal H}^{in}_{int}+{\cal H}^{in}_{light}$, 
 with
\begin{eqnarray}
{\cal H}^{in}_{atom} &=& \sum_{j=1}^3\int dx\;\hat{\psi}^{in\dag}_j(x)H^{in}_j\hat{\psi}^{in}_j(x),\nonumber\\
{\cal H}^{in}_{int} &=& \hbar\int dx\;\left(\hat{\psi}^{in}_2(x)\hat{\psi}^{in\dag}_3(x)\Omega^{in}(x,t)+h.c.\right)\nonumber\\
&&
+\hbar g_{13}\int dx\;\left(\hat{E}^{in}(x)\hat{\psi}^{in}_1(x)\hat{\psi}^{in\dag}_3(x)+h.c.\right),\nonumber\\
{\cal H}^{in}_{light} &=& \int dx\;\hat{E}^{in\dag}(x)pc\hat{E}^{in}(x),
\label{eq:sendhams}
\end{eqnarray}
where $H^{in}_1=-\frac{\hbar^2\partial_x^2}{2m}+V^{in}_{1}(x)$, $H^{in}_2=-\frac{\hbar^2\partial_x^2}{2m}+V^{in}_{2}(x)$, $H^{in}_3=-\frac{\hbar^2\partial_x^2}{2m}+\hbar\omega_0+V^{in}_{3}(x)$, $m$ is the atomic mass, and the $V^{in}_{j}$ represent both linear (trapping for $\hat{\psi}^{in}_{1}$) and nonlinear (scattering) potentials. The optical control field is $\Omega^{in}(x,t)=\Omega^{in}_{23}e^{i(k_0x-(\omega_0-\Delta)t)}$ where $\Omega^{in}_{23}$ is the Rabi frequency for the $|2\rangle \to |3\rangle$ transition and $\omega_0$ is the frequency of the $|3\rangle\to|2\rangle$ transition. $\hat{\psi}^{in}_1(x)$, $\hat{\psi}^{in}_2(x)$, $\hat{\psi}^{in}_3(x)$ and $\hat{E}^{in}(x)$ are the annihilation operators for the trapped condensate mode (internal state $|1\rangle$), input atomic beam ($|2\rangle$), excited state atoms ($|3\rangle$), and probe beam photons respectively, satisfying the usual bosonic commutation relations, $[\hat{\psi}^{in}_i(x),\hat{\psi}^{in\dag}_j(x^\prime)]=\delta_{ij}\delta(x-x^\prime)$ and $[\hat{E}^{in}(x),\hat{E}^{in\dag}(x^\prime)]=\delta(x-x^\prime)$.
The coupling coefficient is $g_{13}=(d_{13}/\hbar)\sqrt{\hbar\omega_k/2\epsilon_0A}$, where $d_{13}$ is the electric dipole moment for the $|1\rangle \to |2\rangle$ transition, $\omega_k=ck$, and $A$ is the cross-section of the atom-light interaction region (we use $A$ corresponding to a control laser waist of $100\mu m$). We neglect interatomic interactions on the basis that the atomic beam is dilute and the process will take place over a time short enough that any phase diffusion effects will be minimal. 
We now introduce the rotating frame fields $\tilde{\psi}^{in}_3(x)=\hat{\psi}^{in}_3(x)e^{i(\omega_0-\Delta)t}$ and $\tilde{E}^{in}(x)=\hat{E}^{in}(x)e^{i(\omega_0-\Delta)t}$ and adiabatically eliminate the weakly occupied intermediate state~\cite{Haine2005,ALE}
$\tilde{\psi}^{in}_3(x)\to-\frac{\Omega_{23}}{\Delta}e^{ik_0x}\hat{\psi}^{in}_2(x)-\frac{g_{13}}{\Delta}\tilde{E}^{in}(x)\hat{\psi}^{in}_1(x)$.
We approximate the highly occupied trapped condensate as a coherent state, $\hat{\psi}^{in}_{1}(x,t)=\phi^{in}(x,t)\equiv\langle\hat{\psi}^{in}_{1}(x,t)\rangle$, while allowing the occupation and the spatial shape to change. This is a very accurate approximation if collisional interactions are small. To simplify notation we set $\hat{\psi}^{in}_2\equiv\hat{\psi}^{in}$ to arrive at the equations of motion
\begin{eqnarray}
i\frac{d \hat{\psi}^{in}}{d t} &=& H^{in}_a\hat{\psi}^{in}(x)-\Omega^{in}_C(x)e^{-ik_0x}\tilde{E}(x),\nonumber\\
i\frac{d\tilde{E}^{in}}{dt} &=& H^{in}_b\tilde{E}^{in}(x)-\Omega_C^*(x)e^{ik_0x}\hat{\psi}^{in}(x),\nonumber\\
i\frac{d\phi^{in}}{dt} &=& H^{in}_{\phi}\phi^{in}(x)-\frac{g_{13}\Omega_{23}}{\Delta}e^{ik_0x}\langle \hat{E}^{in\dag}(x)\hat{\psi}^{in}(x)\rangle ,
\label{eq:eoms}
\end{eqnarray}
with $H^{in}_a=-\hbar\partial_x^2/2m-|\Omega_{23}|^2/\Delta$, $H^{in}_b=-ic\partial_x-|\phi^{in}(x)|^2(g_{13})^2/\Delta+\Delta-\omega_0$,
$H^{in}_{\phi}=-\hbar\partial_x^2/2m+V^{in}_1(x)/\hbar-\langle\hat{E}^{in\dag}(x)\hat{E}^{in}(x)\rangle(g_{13})^2/\Delta$, and 
$\Omega_C(x)=\phi^{in}(x)\Omega_{23}^*g_{13}/\Delta$. The equations of motion for the equivalent variables at the {\em receiver\/} position are found by a similar procedure and essentially differ only in the initial conditions and translation on the $x$-axis.
As shown in Ref.~\cite{Haine2005}, equations of this type can be efficiently solved to give all relevant observables.

\begin{figure}\begin{centering}\includegraphics[width=1\columnwidth]{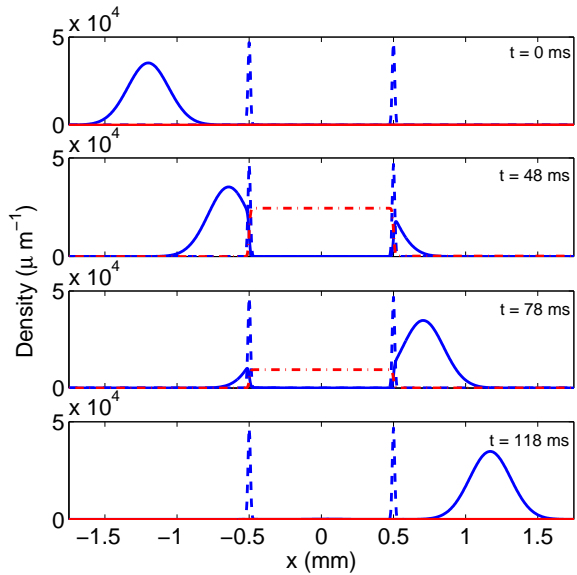}\par\end{centering}
\vspace*{-10pt}
\caption{At the top, the initial atomic pulse (solid curve) is shown about to enter the sending condensate (left dashed curve). The second and third pictures show the pulse partially absorbed, with an output pulse exiting the receiving condensate (right dashed curve). In the bottom picture, the process is completed, with a reconstructed atom laser pulse propagating in free space. The atom pulse and optical probe (dashed-dotted curve) are magnified by factors of $1000$ and $1000\times mc/2\hbar k_0$ to plot them on the condensate scale.}
\label{fig:gokirk}
\vspace*{-10pt}
\end{figure}
The results of our one-dimensional numerical calculations are shown in Fig.~\ref{fig:gokirk}.
We used an input atomic pulse of $n_0=5\times 10^3$, with momentum wavevector $2k_0$, coupled into the initial condensate, where $k_{0}= 8\times 10^6\;$m$^{-1}$, giving an atom laser beam velocity of $v_{atom} = 1.1\;$cm\,s$^{-1}$.  We use $N_0=10^6$ atoms at each site, trapped with potentials $V(x)=m\omega_t^2x^2/2$ and frequencies $\omega_t=5\;$Hz. In all cases we operate at the optimal efficiency point for the signal so that the ratio of the condensate width to the mean beam velocity is tuned to one quarter of a Rabi cycle, $T_{\rm Rabi}\approx 4\sqrt{\hbar/m\omega_t}(m/2\hbar k_0)$. 
The two condensates are shown by the dashed lines, which are given as $1$mm apart, merely for convenience. In the uppermost panel, an atom laser pulse is about to enter the region of the first trapped condensate. The middle two panels show the pulse being incoupled and the probe field, which was initially vacuum in this example, transmitting between the two condensates. The beginnings of the pulse can be seen as the outcoupling process proceeds. In the lowest panel, the outcoupling process has been completed and a replica of the initial pulse is propagating away from the second condensate. Apart from the information needed to prepare the two condensates with near identical numbers and the control fields with similar intensities, no information except that contained in the propagating probe field has been exchanged. There is no requirement for the sharing of Bell pairs or modes of an entangled state as in regular quantum teleportation protocols. In principle, this scheme can be operated with a very high fidelity, as long as the appropriate Rabi frequencies are matched at each site. In practice, this would mean having sending and receiving condensates of comparable sizes and control lasers of equal intensities. Once this was achieved, the remaining sources of degradation are phase diffusion from collisional interactions and spontaneous emission from the excited electronic levels of the two trapped condensates. As shown in previous work, these can be minimised~\cite{homodyne,ALE}. We also note here that the trapped condensates need not be interacting, as although interatomic collisions are necessary for the evaporative cooling which leads to condensation, they are not necessary to maintain the condensed state once a BEC has formed.

\begin{figure}\begin{centering}\includegraphics[width=1\columnwidth]{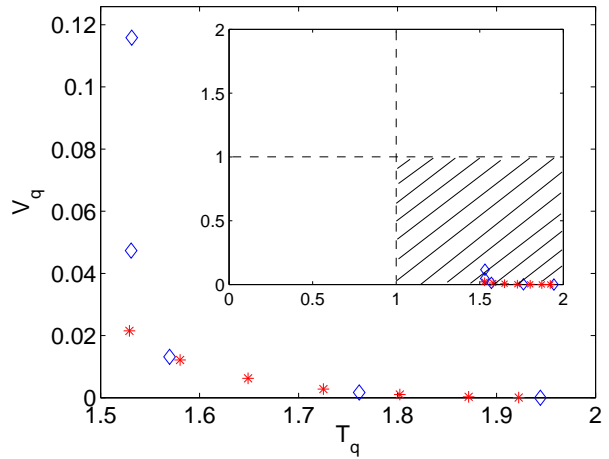}\par\end{centering}
\vspace*{-10pt}
\caption{Teleportation quality on a $T-V$ graph as the ratio $\Omega^{in/out}/\Omega^{in/out}_{opt}$ is varied between $0.66$ and $1.33$ while the atom numbers are equal (diamonds), and the atomic number difference ratio, $\Delta N$ (defined in the text), is varied over the range $-0.66$ to $0.66$ (stars) with fixed Rabi frequency optimised for $\Delta N=0$. The best values (highest $T_{q}$ and lowest $V_{q}$ are for optimal Rabi frequencies and $N_1=N_2$, with the results being symmetric on both sides of these values. The inset shows the complete phase-space, demonstrating that our results are well inside the desired region. As can be seen, the efficiency is more degraded by changes in the control field Rabi frequencies than by changes in the relative numbers of atoms in the sending and receiving condensates.}
\label{fig:yahoo}
\vspace*{-10pt}
\end{figure}
In principle, as long as there are no losses and all the probe light emitted at the first condensate interacts with the second, the fidelity of this process can be $100\%$. In practice there will always be sources of degradation, although they would need to be significant before the efficiency of the process fell to the $2.2\%$ obtained in the Ginsberg experiment~\cite{Ginsberg}. We will now examine the probable effects of the mechanisms which could be responsible for loss of fidelity. 
Apart from the stability of the lasers used, the main sources are spontaneous emission losses from the excited atomic level, phase noise due to atomic collisions, different numbers in the sending and receiving condensates, and non-ideal intensities of the two control lasers. Using our calculational technique, we are able to quantify the last two of these effects, while we can make informed estimates of the first three. 

One method which has previously been used to measure the accuracy of teleportation involves the transfer coefficients as used by Bowen {\em et al.}~\cite{Bowen} for continuous variable quantum teleportation, and is based on quantum nondemolition criteria as developed by Grangier {\em et al.}~\cite{Grangier}. These measures give more information about our system than the conventional measures of fidelity used in standard quantum teleportation, and work well for Gaussian states. The measures are the signal transfer, $T_{q}$, and noise correlation $V_{q}$, which together define the accuracy of transfer of quantum information between the sender and receiver. For our purposes, we note that perfect reconstruction of an input state gives $T_{q}=2$ and $V_{q}=0$, with $T_{q}>1$ violating the information cloning limit and $V_{q}<1$ denoting some degree of quantum reconstruction. In order to calculate these factors, we have used a minimum uncertainty squeezed input pulse with quadrature variances $V(\hat{X})=0.14$ and $V(\hat{Y})= 7.39$, where the $\hat{X}$ and $\hat{Y}$ quadratures are the mode-matched real and imaginary parts of the atomic field~\cite{homodyne}. We compute the $T-V$ diagram by projecting the output state onto a translated copy of the input state.
In Fig.~\ref{fig:yahoo} we show results for the transfer of this pulse, while we vary the control Rabi frequency (diamonds) and the atomic number difference ratio, $\Delta N$, (stars) around the optimal values used to produce Fig.~\ref{fig:gokirk}. $\Delta N$ is defined as $\delta/N$, where $N$ is the fixed total number in the two condensates and the individual numbers are varied as $N_{1}=(N+\delta)/2$ and $N_{2}=(N-\delta)/2$. 
Over the range investigated these coefficients stayed in the desired region of the graph, showing that this technique should be extremely robust to experimental imperfections.

The effect of spontaneous emission can be estimated from the spontaneous emission rate for a transition with frequency $\omega_0=k_0 c$ radiating into a continuum, $\gamma_{\rm sp}=k_0^3|d_{13}|^2/3\pi\hbar\epsilon_0$.
The total spontaneous loss at each station is then
$L_{\rm sp}=\gamma_{\rm sp}\int dx\;\int dt\; \langle \hat{\psi}_3^\dag(x,t)\hat{\psi}_3(x,t)\rangle$.
Using the adiabatically eliminated expression for the excited state $\langle \hat{\psi}_3^\dag(x,t)\hat{\psi}_3(x,t)\rangle\approx\langle \hat{\psi}_2^\dag(x,t)\hat{\psi}_2(x,t)\rangle(\Omega_{23}/\Delta)^{2}$, and the fact that each excited atom on average remains excited for time $T_{\rm Rabi}/4$, we have $L_{\rm sp}\lesssim\gamma_{\rm sp}\bar{N}_3T_{\rm Rabi}/4$, where $\bar{N}_3$ is the total number of excited state atoms transferred.
For the processes to remain coherent, we require the spontaneous emmission per input atom,
$L_{\rm sp}/N_2$, to be small. We find $L_{\rm sp}/N_2\approx 0.04$ at each condensate, allowing us to estimate the effect phenomenologically using a beam splitter which mixes the signals and vacuum with reflectivity $\eta$ ($\approx 0.04$ here), which gives a transfer efficiency of $(1-\eta)^{2}=0.92$ for the whole process. 
The effect of atomic collisions will be greatest within the trapped condensate, but as we are transferring the statistics of the input field to the probe light and back, these will have little effect. The collisions between the propagating and the trapped atoms will have two undesired effects. Firstly, there will be a mean-field effect which will tend to rotate the quadrature phases, but will not affect intensities. The second effect will be that of phase-diffusion of the input pulse, which will only be significant if the pulse is longer than the coherence length. We may consider this effect by noting that the velocity transferred to or from $^{23}$Na by the Raman transition can be up to $6$ cm/s, with up to $1.2$ cm/s for $^{87}$Rb. Using a single-mode expression for the phase diffusion~\cite{Steel1998} and the parameters of our simulations, we find that $^{23}$Na can travel up to $3$mm and $^{87}$Rb up to $600\mu$m in their respective coherence times. As this is larger than the diameter of present condensates, the effect over the coupling times will be small.

In conclusion, we have described a scheme which can be used to imprint the quantum information of an atomic pulse onto an optical field which may then be used to reconstruct an equivalent pulse at a distant location. The process does not depend on entanglement sharing between the sending and receiving stations and the quantum or classical nature depends solely on the state of the atomic field which is transferred. If this is in a classical state, the process will not involve any exotic quantum states and may in some sense be considered as classical teleportation. If the input pulse is squeezed or in some other quantum state, the probe field will also be in this state, so that there will be a transfer of quantum information. 
Our scheme is possible with existing technology and in principle provides for the disembodied optical transmission of a macroscopic matter wave over large distances.

\begin{acknowledgments}

We acknowledge interesting discussions with Crispin Gardiner and Andrew Doherty. Financial support was provided by the Australian Research Council and the Queensland state government.

\end{acknowledgments}

\end{document}